\documentclass[conference, 10pt]{IEEEtran}
\pdfoutput=1
\IEEEoverridecommandlockouts
\usepackage[usenames,dvipsnames]{xcolor}
\usepackage{cite}
\usepackage{epsf}
\usepackage{psfrag}
\usepackage{amsmath,amssymb,amsfonts}
\usepackage{algorithmic}
\usepackage{graphicx}
\usepackage{textcomp}
\usepackage{color}
\usepackage{xcolor}

\def\BibTeX{{\rm B\kern-.05em{\sc i\kern-.025em b}\kern-.08em
		T\kern-.1667em\lower.7ex\hbox{E}\kern-.125emX}}
\newtheorem{theorem}{Theorem}

\newtheorem{lemma}{Lemma}
\newtheorem{definition}{Definition}
\newtheorem{remark}{Remark}

\begin{document}
	
	\title{Remote Empirical Coordination \\
		{\footnotesize \textsuperscript{}}}
	%\begin{figure}
		%\centering
	%	\includegraphics[width=0.7\linewidth]{"Imperfect Empirical Coordination"}
	%	\caption{}
	%	\label{fig:imperfect-empirical-coordination}
	%\end{figure}
	%}
	
	\author{\IEEEauthorblockN{Michail Mylonakis}
		\IEEEauthorblockA{\textit{Division of Inf. Science \& Eng.} \\
			\textit{KTH Royal Institute of Technology}\\
			%Stockholm, Sweden \\
			mmyl@kth.se}
		\and
		\IEEEauthorblockN{ Photios A. Stavrou}
		\IEEEauthorblockA{\textit{Division of Inf. Science \& Eng. } \\
			\textit{KTH Royal Institute of Technology}\\
			%Stockholm, Sweden \\
			fstavrou@kth.se}
		\and
		\IEEEauthorblockN{ Mikael Skoglund}
		\IEEEauthorblockA{\textit{Division of Inf. Science \& Eng.} \\
			\textit{KTH Royal Institute of Technology}\\
			%Stockholm, Sweden \\
			skoglund@kth.se}}
	%\and
	%\and
	%\and
	%\and
	
	\maketitle
	
\begin{abstract}
	We apply the framework of imperfect empirical coordination to a two-node setup where the action $X$ of the first node is not observed directly  but via $L$ agents who observe independently impaired measurements $\hat X$ of the action. These $L$ agents, using a rate-limited communication that is available to all of them, help the second node to generate the action $Y$ in order to establish the desired coordinated behaviour. When $L<\infty$, we prove that it suffices $R_i\geq I\left(\hat X;\hat{Y}\right)$ for at least one agent whereas for $L\longrightarrow\infty$, we show that it suffices $R_i\geq I\left(\hat X;\hat Y|X\right)$ for all agents where $\hat Y$ is a random variable such that $X-\hat X-\hat Y$ and $\|p_{X,\hat Y}\left(x,y\right)-p_{X,Y}\left(x,y\right)\|_{TV}\leq \Delta$ ($\Delta$ is the pre-specified fidelity).

\end{abstract}

\section{Introduction}
The development of machine to machine communication and the Internet of Things has enabled a renewed interest in further investigating heterogeneous network topologies where various objects are allowed to be interconnected. Such objects may be for instance computers with different operating systems and protocols, embedded sensors, medical devices, smart meters, and autonomous vehicles. A key factor to elucidate further insights of such network topologies is to study the cooperation and coordination of the different devices in the network on the
	level of information theory. % this can be established using fundamental ideas  development necessitates the need for distributed coding between the nodes of a network depending on the action taken.}
\par In many practical scenarios, there is no direct access to the source data of some phenomenon due to possible technical limitations. In this case, multiple agents can be deployed to collect noisy measurements of the source. Examples include the {\it remote source coding problem} introduced in \cite{dobrushin:1962} (see also \cite{berger:1971, witsenhausen:1980})	and  the {\it CEO problem} introduced in \cite{berger:1996}. Here, we adopt the concept of the ``remote source'' to the framework of {\it``imperfect'' empirical coordination} \cite{mylonakis:2019} using also ideas from the framework of {\it ``perfect'' empirical coordination} \cite{cuff:2010}.
\par The notion of empirical coordination in information theory was formalized in \cite{cuff:2010}. According to \cite{cuff:2010}, when we are given the actions of some nodes by nature, empirical coordination is achieved if the joint type, measured by total variation distance, of the actions of all nodes in a network is close to the desired distribution, in probability. The literature on empirical coordination is vast. For instance, the authors in \cite{bereyhi:2013,letreust:2015a} studied empirical coordination for various network topologies, whereas in \cite{chou:2018} empirical coordination was established using polar coding and distributed approximation. This type of coordination is also used with ideas from other fields, such as game theory \cite{letreust:2016}, optimal control \cite{letreust:2018} and networked control systems \cite{sharieepoorfard:2018}. The framework of empirical coordination of \cite{cuff:2010} was recently extended to the more general framework of imperfect empirical coordination in \cite{mylonakis:2019} that was inspired by \cite{kramer:2007}. According to \cite{mylonakis:2019}, imperfect empirical coordination is established if the total variation between the joint type of the actions in a network comes close, on average, to a desired distribution within distance pre-specified by a threshold $\Delta$. The choice of $\Delta$ regulates the coordination rates between the agents and therefore the system's designer can choose to coordination in a range of rates depending on the available rate budget.  Clearly, if we choose $\Delta=0$, then, we obtain as a special case the perfect empirical coordination of \cite{cuff:2010}. The result in \cite{mylonakis:2019} was applied to a multiple description problem with two channels in \cite{mylonakis:2019b}. %\cite[Theorem 1]{mylonakis:2019} shows that the knowledge of the capacity region for the problem of perfect empirical coordination gives us immediately the  rate-distortion-coordination region for the problem of imperfect coordination.}
%Clearly, if $\Delta=0$, then, we obtain the empirical coordination in the sense of \cite{cuff:2010}.
\begin{figure}
	\begin{center}
		
		\includegraphics[width=5cm,height=4cm,keepaspectratio]{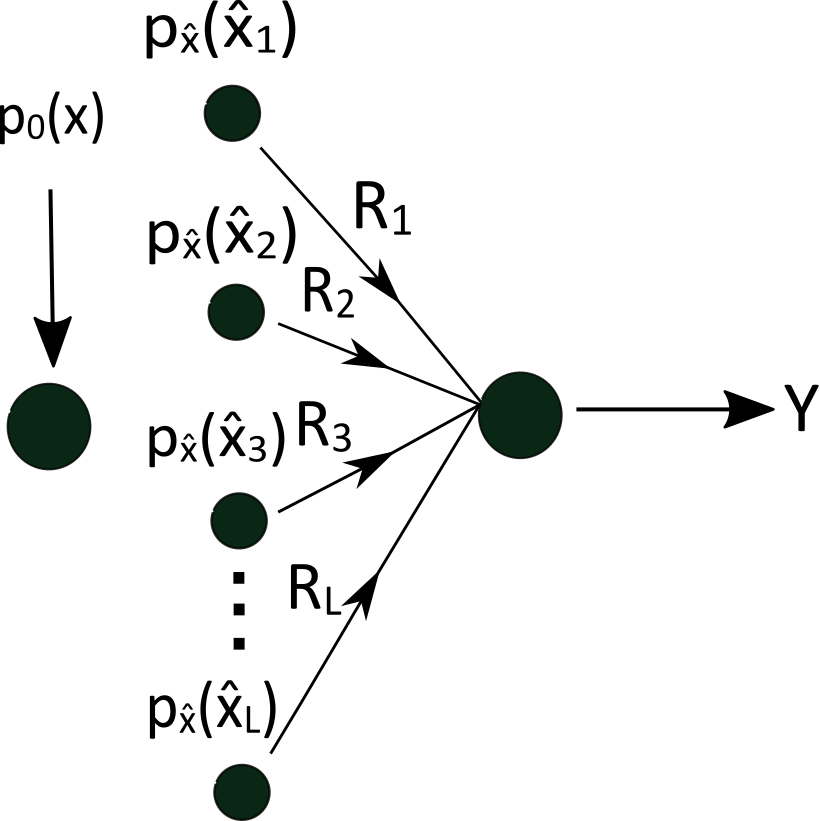}
	\end{center}
	\caption{System model.}
	%	Each node performs an action where some of these actions are selected randomly by nature. 
	\label{fig1}
\end{figure}

In this work, we consider the setup illustrated in Fig. \ref{fig1}. In this setup, the action of the first node, which is distributed according to $p_{0}$, is partially observed via multiple agents who then communicate via multiple rate-limited links to the second node. In particular, the $L$ agents collect independently noisy versions of the action, distributed according to $p_{\hat X}$, and, by applying the coordination code, communicate to the second node. Based on the messages that it receives, the second node produces the action $Y$. Through our framework, we claim that imperfect empirical coordination is an appropriate approach to study coordination of nodes which do not directly communicate. This is because, by definition, the metric to achieve perfect empirical coordination can only be satisfied if the desired distributions satisfy the Markov chain  $X-\hat{X}-Y$ which is not a necessary requirement in imperfect empirical coordination due to the flexibility of our achievability performance criterion. Our achievability results rely on \cite[Theorem 1]{mylonakis:2019} and we break our derivations in two parts. First, in sections III and IV, we give a lower bound of the coordination capacity region for the problem of perfect empirical coordination \cite{cuff:2010}. Second, in section V, we apply \cite[Theorem 1]{mylonakis:2019}  to get a lower bound of the rate-distortion-coordination region. It is noteworthy to point out that our results are obtained for $L<\infty$ and when $L\longrightarrow\infty$. 
     
	%For instance, consider a network where due to packet loss some part of the message does not arrive at the receiver. A traditional coordinated system will fail, while a multiple
	%description system can still coordinate the nodes, with less
	%accuracy.

\section{General Definitions}
We begin with some basic mathematical concepts and the definition of the coordination code i.e., the protocol which is used to coordinate the nodes of the network. We denote as $\mathbb X$ the (common) alphabet of random varibales $X$ and $\hat X$ and as $\mathbb Y$ the (common) alphabet of $Y$ and $\hat Y$.
%In this section, we state the definitions of perfect and imperfect empirical coordination in the context of the cascade network of Fig \ref{fig:fig2}. These definitions have obvious generalizations to other networks. We begin with some basic mathematical concepts and the definition of the $\Delta$-neighborhood, a concept which will help us in the statement of our results.
\begin{definition}[Joint type] The joint type $P_{x^n,y^n}$ of a tuple of sequences $\left(x^n,y^n\right)$ is the empirical probability mass function, given by
	\begin{equation*}P_{x^n,y^n}\left(x,y\right)\triangleq \frac{1}{n}\sum_{i=1}^n{\mathbf 1\big(\left(x_i,y_i\right)=\left(x,y\right)\big)},\end{equation*}
	for all $\left(x,y\right)\in \mathbb{X}\times \mathbb{Y}$, where $\mathbf 1$ is the indicator function.
\end{definition}

\begin{definition}[Total variation]The total variation between two probability mass functions (PMF) is given by \begin{equation*}\|p\left(x,y\right)-q\left(x,y\right)\|_{TV}\triangleq\frac{1}{2}\sum_{x,y}{|p\left(x,y\right)-q\left(x,y\right)|}.\end{equation*}
\end{definition}
	\begin{definition}[$\Delta$-neighborhood]
	The $\Delta$-neighborhood of a PMF $p\left(x,y\right)$ is defined as
	\begin{IEEEeqnarray*}{rCl} N_{\Delta}\big(p\left(x,y\right)\big)\}\triangleq \big\{q(x,y):\|p\left(x,y\right)-q\left(x,y\right)\|_{TV}\leq\Delta\big\}.
	\end{IEEEeqnarray*}
\end{definition}
%\begin{definition}[$\Delta$-neighborhood]
%The $\Delta$-neighborhood of a distribution $p\left(x,y,z\right)$ is defined as
%\begin{IEEEeqnarray*}{rCl}\IEEEeqnarraymulticol{3}{l} {N_{\Delta}\big(p\left(x,y,z\right)\big)}\\&\triangleq& \big\{q(x,y,z):\|p\left(x,y,z\right)-q\left(x,y,z\right)\|_{TV}\leq\Delta\big\}.
%	\end{IEEEeqnarray*}
%\end{definition}

%\subsection{Coordination Code}
%A $\left(2^{nR_1},2^{nR_2},n\right)$ coordination code is the protocol which %is used  to coordinate the actions in the network for a block
%of n time periods. The coordination code and the distribution
%of the random actions, $X_n$, induce a joint distribution on
%the actions in the network.
\begin{definition}[Coordination code]The $\left(2^{nR_1},2^{nR_2},\cdots,2^{nR_L},n\right)$ coordination code for our set-up consists of $L+1$ functions-L encoding functions 
	\begin{equation*}
	i_l:\mathbb X^n%\times\Omega%
	\rightarrow\left\{1,\dots,2^{nR_l}\right\}, l=1,\dots,L,
	\end{equation*}
	and a decoding function
	\begin{equation*}
	y^n:\left\{1,\dots,2^{nR_1}\right\}\times \dots \times \left\{1,\dots,2^{nR_L}\right\}  %
	\rightarrow \mathbb Y^n.
	\end{equation*} 
\end{definition}
%\begin{definition}[Induced distribution\cite{cuff:2010}]
%The induced distribution $\tilde{p}\left(x^n, y^n, z^n\right)$ is the resulting joint distribution of the actions in the network $X^n, Y^n$, and $Z^n$ when a $\left(2^{nR_1},2^{nR_2},n\right)$ coordination code is used.
%\end{definition}

In our set-up, the actions $X^n$ and  ${\hat{X}_l}^n$ for $l=1,\dots,L$ are chosen by nature to be i.i.d according to $p_{X,\hat X_1,\dots,\hat X_L}\left(x,\hat x_1,\dots,\hat x_L\right)=p_0\left(x\right)\prod_{l=1}^{L}{p_{\hat X|X}\left(\hat x_l|x\right)}$. Thus, $X^n$ and  ${\hat{X}_l}^n$ for $l=1,\dots,L$ are distributed according to a product distribution  $\left(X^n,\hat X_1^n,\dots,\hat X_L^n\right)\sim \prod_{i=1}^{n}{p_0\left(x_i\right)\prod_{l=1}^{L}{p_{\hat X|X}\left(\hat x_{li}|x_i\right)}}$.
%Moreover, define $p_{0x}\left(x\right)=\sum_{\hat{x}}{p_0\left(x,\hat{x}\right)}$.
The action $Y^n$ is function of $\hat X_1^n,\dots,\hat X_L^n $ given by   $Y^n=y^n\bigg(i_1\left(\hat X_1^n\right),\dots,i_L\left(\hat X_L^n\right)\bigg)$. 

\section{Finite number of agents}
In this section, we give and discuss an inner bound of the coordination capacity region for the case of finite $L$. We begin with the required definitions.
\begin{definition}[Achievability for perfect coordination and $L$ finite] A desired PMF $p_{X,Y}\left(x,y\right)\triangleq p_{0}\left(x\right)p_{Y|X}\left(y|x\right)$ is achievable for empirical coordination with the rates $\left(R_1,\dots,R_L\right)$ if there exists a sequence of $\Big(2^{nR_1},\cdots,2^{nR_L},n\Big)$  coordination codes such that as $n\to \infty$
	%if for arbitrary $\epsilon_1,\epsilon_2\geq 0$ there exists a sequence of $\Big(e^{n\left(R_1+\epsilon_1\right)},e^{n\left(R_2+\epsilon_2\right)},n\Big)$  coordination codes such that 
	
	%the total variation between the joint type of the actions in the network and the desired distribution goes to zero in probability (under the induced distribution). That is,
	\begin{equation}\|P_{x^n,y^n}\left(x,y,z\right)-p_{0}\left(x\right)p_{Y|X}\left(y|x\right)\|_{TV}\to 0,
	\label{eq:empcord}
	\end{equation}
	in probability.	\label{def:perfect}	
\end{definition}
\begin{definition}[Coordination capacity region for $L$ finite]
	The coordination capacity region $C_{p_{X,\hat X_1,\dots,\hat X_L}}^P$ for the source-agent joint PMF $p_{X,\hat X_1,\dots,\hat X_L}\left(x,\hat x_1,\dots,\hat x_L\right)=p_0\left(x\right)\prod_{l=1}^{L}{p_{\hat X|X}\left(\hat x_l|x\right)}$  is the closure of the set of rate-coordination tuples $\big(R_1,R_2,\dots,R_L,p_{Y|X}\left(y|x\right)\big)$ that are achievable:
	\begin{equation*}
	C_{p_{X,\hat X_1,\dots,\hat X_L}}^P\triangleq
	\mathbf {Cl}\left\{ \,
	\begin{IEEEeqnarraybox}[
	\IEEEeqnarraystrutmode
	\IEEEeqnarraystrutsizeadd{1pt}
	{1pt}][c]{l}
	\big(R_1,\dots,R_L,p_{Y|X}\left(y|x\right)\big):\\
	p_{0}\left(x\right)p_{Y|X}\left(y|x\right)\\\text{is achievable at rates $\left(R_1,\dots,R_L\right)$}
	\end{IEEEeqnarraybox}\right\}.
	\end{equation*}
	%\begin{equation*}C_{p_0}^P\triangleq \mathbf {Cl}\Big\{\big(R_1,R_2,p_{Y,Z|X}\left(y,z|x\right)\big):p_{0}\left(x\right)p_{Y,Z|X}\left(y,z|x\right)\text{is achievable at rates $\left(R_1,R_2\right)$}\Big\}.\end{equation*} 
\end{definition}
\begin{theorem}
The following region is a subset of the coordination capacity region $C_{p_{X,\hat X_1,\dots,\hat X_L}}^P$ for the source-agent joint PMF $p_{X,\hat X_1,\dots,\hat X_L}\left(x,\hat x_1,\dots,\hat x_L\right)=p_0\left(x\right)\prod_{l=1}^{L}{p_{\hat X|X}\left(\hat x_l|x\right)}$:
	\begin{equation*}
C_{p_{X,\hat X_1,\dots,\hat X_L}}^P\supseteq
\left\{ \,
\begin{IEEEeqnarraybox}[
\IEEEeqnarraystrutmode
\IEEEeqnarraystrutsizeadd{1pt}
{1pt}][c]{l}
\big(R_1,\dots,R_L,p_{Y|X}\left(y|x\right)\big):
\quad X-\hat X-Y,\\
\exists l \quad \text{such that} \quad
R_l\geq I\left(\hat X;Y\right)
\end{IEEEeqnarraybox}\right\}.
\end{equation*}
\label{th:fiag}
\end{theorem}
\begin{IEEEproof}
%\textbf{Achievability}.
See Appendix A.
\end{IEEEproof}
\begin{remark} According to Theorem \ref{th:fiag}, in the case of $L< \infty$, the PMFs $p_{X,Y}\left(x,y\right)\triangleq p_{0}\left(x\right)p_{Y|X}\left(y|x\right)$ which form a Markov chain $\quad X-\hat X-Y$, are achievable if the rate of at least one agent exceeds the mutual information between $\hat X$ and $Y$. Although we do not prove an outer bound, it seems to us that, if the number of agents is finite and the rate of all of them is under the thresold of $I\left(\hat X;Y\right)$, the establishment of perfect empirical coordination is impossible i.e., that is optimal to deactivate all but one agent with rate at least equal to $I\left(\hat X;Y\right)$.  On the other hand, as we will see in the next section, if the number of agents is allowed to become arbitrarily large, then, the joint decoding is becoming gainful and we can distribute the rate among the different agents in order to satisy the coordination criterion. 
\end{remark}
\section{Infinite number of agents}
In this section, we give and discuss an inner bound of the coordination capacity region for the case of $L\longrightarrow\infty$. For a proof, see Appendix A. We begin with the required definitions. 
\begin{definition}[Achievability for perfect coordination and $L\to \infty$] A desired PMF $p_{X,Y}\left(x,y\right)\triangleq p_{0}\left(x\right)p_{Y|X}\left(y|x\right)$ is achievable for empirical coordination with the rate per agent $R_{\text{ag}}$ if there exists a sequence of $\Big(\underbrace{2^{nR_{\text{ag}}},\dots,2^{nR_{\text{ag}}}}_{L \quad  \text{times}},n\Big)$  coordination codes such that as $L\to \infty$ and $n\to\infty$ 
	%if for arbitrary $\epsilon_1,\epsilon_2\geq 0$ there exists a sequence of $\Big(e^{n\left(R_1+\epsilon_1\right)},e^{n\left(R_2+\epsilon_2\right)},n\Big)$  coordination codes such that 
	
	%the total variation between the joint type of the actions in the network and the desired distribution goes to zero in probability (under the induced distribution). That is,
	\begin{equation}\|P_{x^n,y^n}\left(x,y,z\right)-p_{0}\left(x\right)p_{Y|X}\left(y|x\right)\|_{TV}\to 0, \label{eq:empcord1}
	\end{equation}
	in probability.	\label{def:perfect}	
\end{definition}
\begin{remark} The double convergence in Definition \ref{def:perfect} should be interpreted as $L\to \infty$ first, followed by $n\to \infty$. See proof of Theorem \ref{th:infiag} (in Appendix A).
\end{remark}
\begin{definition}[Coordination capacity region for $L\to \infty$]
	The coordination capacity region $C_{p_{X,\hat X}}^P$ for the source-agent PMF $p_{X,\hat X}\left(x,\hat x\right)=p_0\left(x\right){p_{\hat X|X}\left(\hat x|x\right)}$ is the closure of the set of rate-coordination tuples $\big(R_{\text{ag}},p_{Y|X}\left(y|x\right)\big)$ that are achievable:
	\begin{equation*}
	C_{p_{X,\hat X}}^P\triangleq
	\mathbf {Cl}\left\{ \,
	\begin{IEEEeqnarraybox}[
	\IEEEeqnarraystrutmode
	\IEEEeqnarraystrutsizeadd{1pt}
	{1pt}][c]{l}
	\big(R_{\text{ag}},p_{Y|X}\left(y|x\right)\big):
	p_{0}\left(x\right)p_{Y|X}\left(y|x\right)\\\text{is achievable at rate per agent $R_{\text{ag}}$}
	\end{IEEEeqnarraybox}\right\}.
	\end{equation*}
	%\begin{equation*}C_{p_0}^P\triangleq \mathbf {Cl}\Big\{\big(R_1,R_2,p_{Y,Z|X}\left(y,z|x\right)\big):p_{0}\left(x\right)p_{Y,Z|X}\left(y,z|x\right)\text{is achievable at rates $\left(R_1,R_2\right)$}\Big\}.\end{equation*} 
\end{definition}
\begin{theorem}
The following region is a subset of the coordination capacity region $C_{p_{X,\hat X}}^P$ for the source-agent PMF $p_{X,\hat X}\left(x,\hat x\right)=p_0\left(x\right){p_{\hat X|X}\left(\hat x|x\right)}$:	
	\begin{equation*}
C_{p_{X,\hat X}}^P\supseteq
\mathbf {Cl}\left\{ \,
\begin{IEEEeqnarraybox}[
\IEEEeqnarraystrutmode
\IEEEeqnarraystrutsizeadd{1pt}
{1pt}][c]{l}
\big(R_{\text{ag}},p_{Y|X}\left(y|x\right)\big):
X-\hat X-Y,\\\quad \quad \quad \quad \quad \quad \quad \quad 
R_{\text{ag}}\geq I\left(\hat X;Y|X\right)
\end{IEEEeqnarraybox}\right\}.
\end{equation*}	
\label{th:infiag}
\end{theorem}
\begin{IEEEproof} 
See Appendix A.
%\textbf{Achievability}.

    % where $C_a$ is the event that  $\left(\mathbf x, \hat {\mathbf x_l}\right) \in \mathbb{A}_{\epsilon^\prime}^{\ast\left(n\right)}\big(p_{X,\hat X}\big)$ for every $l$, all encoders do not declare an error but the decoder finds  more than one $\big(v^{\left(1\right)},\dots,v^{\left(L\right)}\big)$ and  $C_b$ is the event that  $\left(\mathbf x, \hat {\mathbf x_l}\right) \in \mathbb{A}_{\epsilon^\prime}^{\ast\left(n\right)}\big(p_{X,\hat X}\big)$ for every $l$, all encoders do not declare an error but the decoder does not find any $\big(v^{\left(1\right)},\dots,v^{\left(L\right)}\big)$.

	 %$\Big(\left(w^{\left(1\right)},v^{\left(1\right)}\right),\dots,\left(w^{\left(L\right)},v^{\left(L\right)}\right)\Big)$ $\mathbf Y^{\left(l\right)}\left(w^{\left(l\right)}\right)$ for one from the $w^{\left(l\right)}$ that it receives. 
%\end{itemize}
%\textbf{Converse}.	
\end{IEEEproof}
\begin{remark} According to Theorem \ref{th:infiag}, in the case of $L\to \infty$, the PMFs $p_{X,Y}\left(x,y\right)\triangleq p_{0}\left(x\right)p_{Y|X}\left(y|x\right)$ which form a Markov chain $\quad X-\hat X-Y$, are achievable if every agent has rate at least equal to $ I\left(\hat X;Y|X\right)$, which of course is smaller or equal to $I\left(\hat X; Y\right)$ due to the markovian property. In other words, the arbitrarily large number of agents allows us to get rid of the constraint $R_i\geq I\left(\hat X;Y\right)$ for at least one agent.  \end{remark}

\section{Imperfect empirical coordination}
In this section, we combine the inner bounds from the previous two sections with \cite[Theorem 1]{mylonakis:2019} in order to get inner bounds for the rate-distortion-coordination region, both in the cases of $L$ finite and $L\longrightarrow\infty$. 
\subsection{Finite number of agents}
\begin{definition}[Achievability for  $\Delta$-empirical  coordination and $L$ finite] A desired PMF $p_{X,Y}\left(x,y\right)\triangleq p_{0}\left(x\right)p_{Y|X}\left(y|x\right)$ is achievable for $\Delta$-empirical coordination with the rate-pair $\left(R_1,\dots,R_L\right)$ if there is an $N$ such that for all $n>N$, there exists a coordination code $\Big(2^{nR_1},\dots,2^{nR_L},n\Big)$ such that
	%if for arbitrary $\epsilon \geq 0$ and large enough $n$, there exists a coordination code $\Big(2^{nR_1},2^{nR_2},n\Big)$ such that
	\begin{equation*}
	%\uplim_{n \to \infty}
	\mathbb{E}\big\{\|P_{x^n,y^n}\left(x,y\right)-p_{0}\left(x\right)p_{Y|X}\left(y|x\right)\|_{TV}\big\}\leq \Delta.\end{equation*}\label{def:imperfect}
\end{definition}

\begin{definition}[Rate-distortion-coordination region for $L$ finite]
	The rate-distortion-coordination region $R_{p_{X,\hat X_1,\dots,\hat X_L}}^I$ for the source-agent PMF $p_{X,\hat X_1,\dots,\hat X_L}\left(x,\hat x_1,\dots,\hat x_L\right)=p_0\left(x\right)\prod_{l=1}^{L}{p_{\hat X|X}\left(\hat x_l|x\right)}$  and for a fixed conditional distribution $p_{Y|X}\left(y|x\right)$ is defined as:
	
	\begin{multline*}
	R_{p_{X,\hat X_1,\dots,\hat X_L}}^I\big(p_{Y|X}\left(y|x\right)\big)\\\quad \triangleq
	\mathbf{Cl}\left.
	\left\{ \,
	\begin{IEEEeqnarraybox}[
	\IEEEeqnarraystrutmode
	\IEEEeqnarraystrutsizeadd{1pt}
	{1pt}][c]{l}
	\left(R_1,\dots,R_L,\Delta\right):
	p_{0}\left(x\right)p_{Y|X}\left(y|x\right)\\ \text{is achievable for $\Delta$-empirical coordination}\\\text{ at rates} \left(R_1,\dots,R_L\right)
	\end{IEEEeqnarraybox}\right\}.
	\right.
	\end{multline*}

\end{definition}
\begin{lemma}
 For every source-agent PMF $p_{X,\hat X_1,\dots,\hat X_L}\left(x,\hat x_1,\dots,\hat x_L\right)=p_0\left(x\right)\prod_{l=1}^{L}{p_{\hat X|X}\left(\hat x_l|x\right)}$ and for every fixed conditional PMF $p_{Y|X}\left(y|x\right)$:	
	
	\begin{multline*}
 R_{p_{X,\hat X_1,\dots,\hat X_L}}^I\big(p_{Y|X}\left(y|x\right)\big)\\\quad \supseteq
	\left.
	\left\{ \,
	\begin{IEEEeqnarraybox}[
	\IEEEeqnarraystrutmode
	\IEEEeqnarraystrutsizeadd{2pt}
	{2pt}][c]{l}
	\left(R_1,\dots, R_L,\Delta\right):\\
	\left(R_1,\dots, R_L,q_{\hat{Y}|X}\right)\in C_{p_{X,\hat{X}_1,\dots,\hat{X}_L}}^P\quad\\ \text{for some $\hat Y$ which satisfy}\\ p_0\left(x\right)q_{\hat{Y}|X}\left(y|x\right)\in N_{\Delta}\big(p_0\left(x\right)p_{Y|X}\left(y|x\right)\big)\\\text{if such an $\hat Y$ exists}
	\end{IEEEeqnarraybox}\right\}.
	\right.
	%\label{eq:example_left_right2}.
	\end{multline*}
	\label{th:maintheorem}	
\end{lemma}
\begin{IEEEproof} This lemma is a direct consequence of a more general result which is explained and proved in \cite{mylonakis:2019}. See also, Fig. \ref{fig2}.
\end{IEEEproof}
\begin{theorem}
	The following region is a subset of the rate-distortion-coordination region $R_{p_{X,\hat X_1,\dots,\hat X_L}}^I$ for the source-agent PMF $p_{X,\hat X_1,\dots,\hat X_L}\left(x,\hat x_1,\dots,\hat x_L\right)=p_0\left(x\right)\prod_{l=1}^{L}{p_{\hat X|X}\left(\hat x_l|x\right)}$  and for a fixed conditional distribution $p_{Y|X}\left(y|x\right)$  :
	\begin{multline*}
	R_{p_{X,\hat X_1,\dots,\hat X_L}}^I\big(p_{Y|X}\left(y|x\right)\big)\\\quad\supseteq
	\left.
	\left\{ \,
	\begin{IEEEeqnarraybox}[
	\IEEEeqnarraystrutmode
	\IEEEeqnarraystrutsizeadd{2pt}
	{2pt}][c]{l}
	\left(R_1,\dots, R_L,\Delta\right):\\ \exists l \quad \text{such that} \quad
	R_l\geq I\left(\hat X;\hat Y\right)
	\\ \text{for some $\hat Y$ which satisfy} \quad X-\hat X-\hat Y\\ \text{and} \quad  p_0\left(x\right)q_{\hat{Y}|X}\left(y|x\right)\in N_{\Delta}\big(p_0\left(x\right)p_{Y|X}\left(y|x\right)\big)
	\end{IEEEeqnarraybox}\right\}.
	\right.
	%\label{eq:example_left_right2}.
	\end{multline*}
\end{theorem}
\begin{IEEEproof}From Theorem \ref{th:fiag} and Lemma \ref{th:maintheorem}, we obtain the characterization of the theorem.
\end{IEEEproof}
%\begin{remark}For $L=1$, the previous theorem together with a simple converse give \cite[Theorem 1]{kramer:2007}.
%\end{remark}
\subsection{Infinite number of agents}
\begin{definition}[Achievability for  $\Delta$-empirical  coordination and $L\to \infty$] A desired PMF $p_{X,Y}\left(x,y\right)\triangleq p_{0}\left(x\right)p_{Y|X}\left(y|x\right)$ is achievable for $\Delta$-empirical coordination with the rate per agent  $R_{\text{ag}}$ if there is an $\bar{L}$ and an $N$ such that for all $L>\bar{L}$ and $n>N$, there exists a coordination code $\Big(\underbrace{2^{nR_{\text{ag}}},\dots,2^{nR_{\text{ag}}}}_{L \quad \text{times}},n\Big)$ such that
	%if for arbitrary $\epsilon \geq 0$ and large enough $n$, there exists a coordination code $\Big(2^{nR_1},2^{nR_2},n\Big)$ such that
	\begin{equation*}
	%\uplim_{n \to \infty}
	\mathbb{E}\big\{\|P_{x^n,y^n}\left(x,y\right)-p_{0}\left(x\right)p_{Y|X}\left(y|x\right)\|_{TV}\big\}\leq \Delta.\end{equation*}\label{def:imperfect}
\end{definition}
\newpage

\begin{figure}
	\begin{center}
		
		\includegraphics[width=7cm,height=6cm,keepaspectratio]{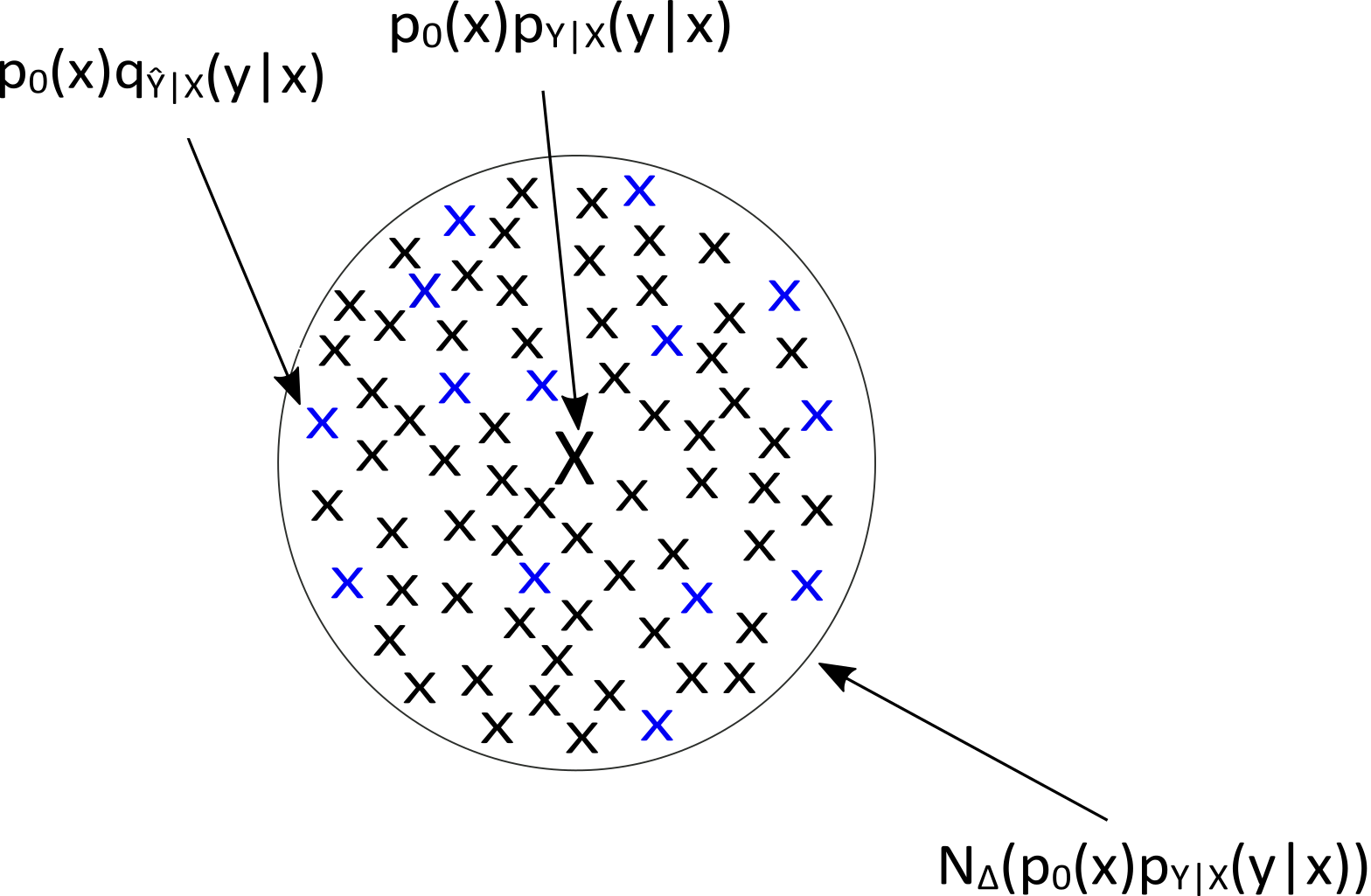}
	\end{center}
	\caption{Interpretation of Lemmas \ref{th:maintheorem} and \ref{th:maintheorem1}: Every good coordination code designed for achieving perfect empirical coordination according to some distribution  $p_0\left(x\right)q_{\hat{Y}|X}\left(y|x\right)\in N_{\Delta}\big(p_0\left(x\right)p_{Y|X}\left(y|x\right)\big)\cap C^P$ (blue) achieves $\Delta$-empirical coordination according to $p_0\left(x\right)p_{Y|X}\left(y|x\right)$.}
	%	Each node performs an action where some of these actions are selected randomly by nature. 
	\label{fig2}
\end{figure}

%\begin{figure}
%	\begin{center}
		
	%	\includegraphics[width=6cm,height=6cm,keepaspectratio]{noisy4.png}
%	\end{center}
%	\caption{Interpretation of Lemmas \ref{th:maintheorem} and \ref{th:maintheorem1} - Converse part: For every coordination code that achieves $\Delta$-empirical coordination according to $p_0\left(x\right)p_{Y|X}\left(y|x\right)$, there is a coordination code with the same rates which achieves perfect empirical coordination according to some distribution in $N_{\Delta}\big(p_0\left(x\right)p_{Y|X}\left(y|x\right)\big)\cap C^P$ (purple).}
	%	Each node performs an action where some of these actions are selected randomly by nature. 
	\label{fig3}
%\end{figure}

\begin{definition}[Rate-distortion-coordination region for $L\to \infty$]
	The rate-distortion-coordination region $R_{p_{X,\hat X}}^I$ for the source-agent PMF $p_{X,\hat X}\left(x,\hat x\right)=p_0\left(x\right){p_{\hat X|X}\left(\hat x|x\right)}$ and for a fixed conditional PMF $p_{Y|X}\left(y|x\right)$ is defined as:
	
	\begin{multline*}
	R_{p_{X,\hat X}}^I\big(p_{Y|X}\left(y|x\right)\big) \triangleq\\
	\mathbf{Cl}\left.
	\left\{ \,
	\begin{IEEEeqnarraybox}[
	\IEEEeqnarraystrutmode
	\IEEEeqnarraystrutsizeadd{1pt}
	{1pt}][c]{l}
	\left(R_{\text{ag}},\Delta\right):
	p_{0}\left(x\right)p_{Y|X}\left(y|x\right) \text{is achievable}\\\text{for $\Delta$-empirical coordination}\text{ at rate per agent $R_{\text{ag}}$}
	\end{IEEEeqnarraybox}\right\}.
	\right.
	\end{multline*}
	
\end{definition}
\begin{lemma}
For every source-agent PMF  $p_{X,\hat X}\left(x,\hat x\right)=p_0\left(x\right){p_{\hat X|X}\left(\hat x|x\right)}$ and for every fixed conditional PMF $p_{Y|X}\left(y|x\right)$,	
	
	\begin{multline*}
 R_{p_{X,\hat X}}^I\big(p_{Y|X}\left(y|x\right)\big)\\\quad=
	\left.
	\left\{ \,
	\begin{IEEEeqnarraybox}[
	\IEEEeqnarraystrutmode
	\IEEEeqnarraystrutsizeadd{2pt}
	{2pt}][c]{l}
	\left(R_{\text{ag}},\Delta\right):\\
	\left(R_{\text{ag}},q_{\hat{Y}|X}\right)\in C_{p_{X,\hat{X}}}^P\quad\\ \text{for some $\hat Y$ which satisfy}\\ p_0\left(x\right)q_{\hat{Y}|X}\left(y|x\right)\in N_{\Delta}\big(p_0\left(x\right)p_{Y|X}\left(y|x\right)\big)
	\end{IEEEeqnarraybox}\right\}.
	\right.
	%\label{eq:example_left_right2}.
	\end{multline*}
	\label{th:maintheorem1}	
\end{lemma}
\begin{IEEEproof} This lemma is a direct consequence of a more general result which is explained and proved in \cite{mylonakis:2019}. See also, Fig. \ref{fig2}.
\end{IEEEproof}

	\begin{IEEEproof} From Theorem \ref{th:infiag} and Lemma \ref{th:maintheorem1}, we obtain the characterization of the theorem.
	\end{IEEEproof}
	%where
	%\begin{multline*}
	%\begin{multline*}
	%R\left(p_0 p_{\hat{Y}|X}\right) \triangleq
	%\left.
	%\left\{ \,
	%\begin{IEEEeqnarraybox}[
	%\IEEEeqnarraystrutmode
	%\IEEEeqnarraystrutsizeadd{1pt}
	%{1pt}][c]{l}
	%\left(R_1,\dots,R_L,\Delta\right):\\	
	%\exists i \quad \text{such that} \quad
	%R_i\geq I\left(\hat X;\hat Y\right)
	%p_0\left(x\right)p_{\hat{Y}^{\left(1\right)}|X}\left(y|x\right)\in N_{\Delta_1},\\
	%p_0\left(x\right)p_{\hat{Y}^{\left(2\right)}|X}\left(y|x\right)\in N_{\Delta_2},\\
	%p_0\left(x\right)p_{\hat{Y}^{\left(12\right)}|X}\left(y|x\right)\in N_{\Delta_{12}}
	%\Delta_1\geq \|p_0\left(x\right)p_{Y|X}\left(y|x\right)-p_0\left(x\right)p_{Y^{\left(1\right)}|X}\left(y|x\right)\|_{TV},\\
	%\Delta_2\geq \|p_0\left(x\right)p_{Y|X}\left(y|x\right)-p_0\left(x\right)p_{Y^{\left(2\right)}|X}\left(y|x\right)\|_{TV},\\
	%\Delta_{12}\geq \|p_0\left(x\right)p_{Y|X}\left(y|x\right)-p_0\left(x\right)p_{Y^{\left(12\right)}|X}\left(y|x\right)\|_{TV}\\
	%\end{IEEEeqnarraybox}\right\}.
	%\right.
	%\end{multline*}
	
\begin{theorem}
The following region is a subset of the rate-distortion-coordination region $R_{p_{X,\hat X}}^I$ for the source-agent PMF $p_{X,\hat X}\left(x,\hat x\right)=p_0\left(x\right){p_{\hat X|X}\left(\hat x|x\right)}$ and for a fixed conditional PMF $p_{Y|X}\left(y|x\right)$:

	\begin{multline*}
	R_{p_{X,\hat X}}^I\big(p_{Y|X}\left(y|x\right)\big)\\\quad\supseteq
	\left.
	\left\{ \,
	\begin{IEEEeqnarraybox}[
	\IEEEeqnarraystrutmode
	\IEEEeqnarraystrutsizeadd{2pt}
	{2pt}][c]{l}
	\left(R_{\text{ag}},\Delta\right):\\
	R_{\text{ag}} \geq \min_{\substack{q_{\hat{Y}|X}:X-\hat{X}-\hat{Y},\\ p_0\left(x\right)q_{\hat{Y}|X}\left(y|x\right)\in N_{\Delta}\left(p_{X,Y}\left(x,y\right)\right)
	}} I\left(\hat X;\hat Y|X\right)
	\end{IEEEeqnarraybox}\right\}.
	\right.
	%\label{eq:example_left_right2}.
	\end{multline*}
\end{theorem}

\section* {\quad \quad \quad \quad Appendix A \newline Proofs of Theorem 1 and Theorem 2}

\begin{IEEEproof}[Proof of Theorem 1]
\begin{itemize}
	\item{\textit{Setup}:} We assume that $\epsilon_l>0$ are given for every $l$. We fix some rates $R_l$, some blocklength $n$, and some $\epsilon>0$
	%$\epsilon_1,\dots,\epsilon_L>0$ 
	and for every PMF $p_{X,\hat X,Y}=p_Xp_{\hat X|X}p_{Y|\hat X}$ compute the marginal $p_{Y}$.	\item{\textit{Codebook design}:} Generate $\lfloor e^{n\left(R_l+\epsilon_l\right)} \rfloor$ length-$n$ codewords $\mathbf Y^{\left(l\right)}\left(w^{\left(l\right)}\right), w^{\left(l\right)}=1,\dots, \lfloor e^{n\left(R_l+\epsilon_l\right)} \rfloor$, by choosing each of the $n\lfloor e^{n\left(R_l+\epsilon_l\right)} \rfloor$ symbols  $Y^{\left(l\right)}_k\left(w^{\left(l\right)}\right), \quad k=1,\dots,n$ independently at random according to $p_{Y}$ for $l=1,\dots,L$.
	\item{\textit{Encoder Design}:} For given sequences $\mathbf{x},\mathbf{\hat x_1},\dots,\mathbf{\hat x_L}$, the $l$-th encoder tries to find a $w^{\left(l\right)}$ such that 
	\begin{multline}
	\bigg( \hat {\mathbf x}_l,\mathbf Y^{\left(l\right)}\left(w^{\left(l\right)}\right)\bigg)
	\in \mathbb{A}_{\epsilon}^{\ast\left(n\right)}\big(p_{\hat{X},Y}\big).\label{eq:jtypicality}
	\end{multline}
	If it finds several possible choices, they pick the first. If it finds none, it declares an error. The $l$-th encoder puts out $w^{\left(l\right)}$. Name $\mathbb L$ the set of indices $l$ for which the $l$-th encoder does not declare an error.
	\item{\textit{Decoder Design}:}  The decoder $y^n$  puts out $\mathbf Y^{\left(l\right)}\left(w^{\left(l\right)}\right)$ for some $l\in \mathbb L$.
	% from the $w^{\left(l\right)}$ that it receives. 
	\item{\textit{Performance Analysis}:} We define $\epsilon^\prime=\frac{\epsilon}{2|\mathbb X|}$ %We define $\epsilon^\prime\triangleq\frac{\epsilon}{2|\mathbb Y|^3}$
	and partition the error space into three disjoint cases:
	%The source sequence $\mathbf x$ is not typical:
	(a) $\left(\mathbf x, \hat {\mathbf x}_l\right) \notin \mathbb{A}_{\epsilon^\prime}^{\ast\left(n\right)}\big(p_{X,\hat X}\big)$ for some $l$
	%in which case we for sure cannot find a pair $\left(w_1,w_2\right)$ such that \eqref{jtypicality} is satisfied
	(b) $\left(\mathbf x, \hat {\mathbf x}_l\right) \in \mathbb{A}_{\epsilon^\prime}^{\ast\left(n\right)}\big(p_{X,\hat X}\big)$ for every $l$ but $\mathbb L$ is empty
	%but every encoder declares an error %i.e. \eqref{jtypicality} is not satisfied for every l=1,\dots,L.
	(c)  $\left(\mathbf x, \hat {\mathbf x}_l\right) \in \mathbb{A}_{\epsilon^\prime}^{\ast\left(n\right)}\big(p_{X,\hat X}\big)$ for every $l$, $\mathbb L$ is not empty
	%and at least one encoder does not declare an error 
	but \eqref{eq:empcord} is not satisfied.
	By the Union Bound and Lemma \ref{lem:TA} (in Appendix B), we can bound the probability of Case (a) as $\Pr\left(\text{Case a}\right)\leq \sum_{l=1}^{L}\delta_{t}\left(n,\epsilon^\prime,\mathbb{X\times X}\right)$.
	In case (b), we get
	\begin{IEEEeqnarray}{rCl}&&
		\Pr\left(\text {Case b}\right)\leq \prod_{l=1}^{L} \Pr\Bigg(\left\{\left(\mathbf x, \hat {\mathbf x}_l\right) \in \mathbb{A}_{\epsilon^\prime}^{\ast\left(n\right)}\big(p_{X,\hat X}\big)\right\}\nonumber\\&&\quad \quad \quad \cap \bigg\{\nexists w^{\left(l\right)}: \Big(\hat {\mathbf x}_l,\mathbf Y^{\left(l\right)}\big(w^{\left(l\right)}\big)\Big)\in \mathbb{A}_{\epsilon^\prime}^{\ast\left(n\right)}\big(p_{\hat X,Y}\big)  \bigg\}\Bigg)\nonumber\\
		&&= \prod_{l=1}^{L}\underbrace{\Pr\Big(\left(\mathbf x, \hat {\mathbf x}_l\right) \in \mathbb{A}_{\epsilon^\prime}^{\ast\left(n\right)}\big(p_{X,\hat X}\big)\Big)}_{\leq 1}\nonumber\\&&\quad \quad \quad \cdot \Pr \bigg( \nexists w^{\left(l\right)}: \Big(\hat {\mathbf x}_l,\mathbf Y^{\left(l\right)}\big(w^{\left(l\right)}\big)\Big)\in \mathbb{A}_{\epsilon^\prime}^{\ast\left(n\right)}\big(p_{\hat X,Y}\big)\nonumber \\&& \quad \quad \quad \quad \quad \Big|\left(\mathbf x, \hat {\mathbf x}_l\right) \in \mathbb{A}_{\epsilon^\prime}^{\ast\left(n\right)}\big(p_{X,\hat X}\big)\bigg)\nonumber\end{IEEEeqnarray}\begin{IEEEeqnarray}{rCl}
	&& \leq \prod_{l=1}^{L}\prod_{w^{\left(l\right)}=1}^{\lfloor e^{n\left(R_l+\epsilon_l\right)} \rfloor} \Pr \bigg( \Big(\hat {\mathbf x_l},\mathbf Y^{\left(l\right)}\big(w^{\left(l\right)}\big)\Big)\notin\mathbb{A}_{\epsilon^\prime}^{\ast\left(n\right)}\big(p_{\hat X,Y}\big)\nonumber\\&&\quad \quad  \quad \quad \quad \quad  \quad \quad \quad \quad  \Big|\left(\mathbf x, \hat {\mathbf x}_l\right) \in \mathbb{A}_{\epsilon^\prime}^{\ast\left(n\right)}\big(p_{X,\hat X}\big)\bigg)\nonumber\\&& = \prod_{l=1}^{L}\prod_{w^{\left(l\right)}=1}^{\lfloor e^{n\left(R_l+\epsilon_l\right)} \rfloor} \Pr \bigg( \Big(\hat {\mathbf x}_l,\mathbf Y^{\left(l\right)}\big(w^{\left(l\right)}\big)\Big)\notin\mathbb{A}_{\epsilon^\prime}^{\ast\left(n\right)}\big(p_{\hat X,Y}\big)\nonumber\\&& \quad \quad  \Big|\left\{ \hat {\mathbf x}_l \in \mathbb{A}_{\epsilon^\prime}^{\ast\left(n\right)}\big(p_{\hat X}\big)\right\}\cap  \left\{\mathbf x\in \mathbb{A}_{\epsilon}^{\ast\left(n\right)}\big(p_{X,\hat X}|\hat{\mathbf x}_l\big)\right\}\bigg)\label{caseb1}\IEEEeqnarraynumspace\\&& = \prod_{l=1}^{L}\prod_{w^{\left(l\right)}=1}^{\lfloor e^{n\left(R_l+\epsilon_l\right)} \rfloor}   \Pr \bigg( \mathbf Y^{\left(l\right)}\big(w^{\left(l\right)}\big)\notin\mathbb{A}_{\epsilon^\prime}^{\ast\left(n\right)}\big(p_{\hat X,Y}|\mathbf x_l\big)\nonumber\\&&\quad \quad \quad \quad  \quad \quad \quad \quad \quad  \Big| \hat {\mathbf x}_l \in \mathbb{A}_{\epsilon^\prime}^{\ast\left(n\right)}\big(p_{\hat X}\big)\bigg)\label{caseb2}\\
		%&& \leq \prod_{l=1}^{L}\prod_{w^{\left(l\right)}=1}^{\lfloor e^{n\left(R_l+\epsilon_l\right)} \rfloor}   1-\Pr \bigg( \mathbf Y^{\left(l\right)}\big(w^{\left(l\right)}\big)\in\mathbb{A}_{\epsilon^\prime}^{\ast\left(n\right)}\big(p_{\hat X,Y}|\hat{\mathbf x}_l\big)\nonumber\\&&\quad \quad \quad \quad  \quad \quad \quad \quad \quad  \Big| \hat {\mathbf x_l} \in \mathbb{A}_{\epsilon^\prime}^{\ast\left(n\right)}\big(p_{\hat X}\big)\bigg)\nonumber
		%&&=\prod_{l=1}^{L}\prod_{w^{\left(l\right)}=1}^{\lfloor e^{n\left(R_l+\epsilon_l\right)} \rfloor}  1-\Pr \bigg( \mathbf Y^{\left(l\right)}\big(w^{\left(l\right)}\big)\in\mathbb{A}_{\epsilon^\prime}^{\ast\left(n\right)}\big(p_{\hat X,Y}|\hat{\mathbf x}_l\big)\nonumber\\&& \Big|\left\{ \hat {\mathbf x_l} \in \mathbb{A}_{\epsilon^\prime}^{\ast\left(n\right)}\big(p_{\hat X}\big)\right\}\cap  \left\{\mathbf x\in \mathbb{A}_{\epsilon}^{\ast\left(n\right)}\big(p_{X,\hat X}|\hat{\mathbf x}_l\big)\right\}\bigg)\label{caseb2}\\&&=\prod_{l=1}^{L}\prod_{w^{\left(l\right)}=1}^{\lfloor e^{n\left(R_l+\epsilon_l\right)} \rfloor}  1-\Pr \Big( \mathbf Y^{\left(l\right)}\big(w^{\left(l\right)}\big)\in\mathbb{A}_{\epsilon^\prime}^{\ast\left(n\right)}\big(p_{\hat X,Y}|\hat{\mathbf x}_l\big)\nonumber\\&&\quad \quad \quad \quad  \quad \quad  \quad \quad \quad \quad \quad  \big| \hat {\mathbf x_l} \in \mathbb{A}_{\epsilon^\prime}^{\ast\left(n\right)}\big(p_{\hat X}\big)\Big)\label{caseb3}\\
		&&=\prod_{l=1}^{L}\prod_{w^{\left(l\right)}=1}^{\lfloor e^{n\left(R_l+\epsilon_l\right)} \rfloor}\bigg( 1-\nonumber\\&&\quad \quad \quad  \Big(1-\delta_t\big(n,\epsilon^\prime/2,\mathbb X\times \mathbb Y\big)\Big) e^{-n\big(I\left(\hat X;Y\right)+2\epsilon_m\big)}\bigg)\label{caseb3} \\&&=\prod_{l=1}^{L}\Bigg(1-\nonumber\\&&\bigg(\Big(1-\delta_t\big(n,\epsilon^\prime/2,\mathbb X\times \mathbb Y\big)\Big) e^{-n\big(I\left(\hat X;Y\right)+2\epsilon_m\big)}\bigg)\Bigg)^{\lfloor e^{n\left(R_l+\epsilon_l\right)}\rfloor}\label{caseb4}\\&&=\prod_{l=1}^{L}\exp\bigg(-\lfloor e^{n\left(R_l+\epsilon_l\right)}\rfloor\nonumber\\&&  \quad \quad \quad \quad \cdot\Big(1-\delta_t\big(n,\epsilon^\prime/2,\mathbb X\times \mathbb Y\big)\Big) e^{-n\big(I\left(\hat X;Y\right)+2\epsilon_m\big)}\bigg)\label{caseb5} \IEEEeqnarraynumspace\\&&=\prod_{l=1}^{L}\exp\Big(-e^{n\left(R_l-I\left(\hat X;Y\right)+\epsilon_l-\delta_l\right)}\Big),\nonumber
		%\begin{IEEEeqnarray*}{r}
		% \Pr\left(\text {Case b}\right)\leq \exp\Big(-\sum_{l=1}^L e^{n\left(R_l-I\left(\hat X;Y\right)+\epsilon_l-\delta_l\right)}\Big)
	\end{IEEEeqnarray}
	where $\delta_l$ accounts for the rounding mistake and includes the $2\epsilon_m$-term and the $\left(1-\delta_t\right)$-factor. So, we see that as long as $n$ is large enough, $R_l\geq I\left(\hat X;Y\right)$ for at least one $l$ and $\epsilon$ small enough such that $\delta_l<\epsilon_l$ for this $l$, the probability $\Pr\left(\text {Case b}\right)$ tends to zero double-exponentially fast in $n$. Here, \eqref{caseb1} results from Lemma \ref{lem:chtypsets} (in Appendix B), \eqref{caseb2} follows again from Lemma \ref{lem:chtypsets} (in Appendix B) and because we discard irrelevant information, \eqref{caseb3}  follows from Lemma \ref{lem:TC} (in Appendix B), \eqref{caseb4} holds because the factor in the product does not depend on $w$ anymore and \eqref{caseb5} follows from Lemma \ref{lem:expineq} (in Appendix B). In case (c),
	\begin{IEEEeqnarray}{rCl}
		&&\Pr\left(\text{Case c}\right)\nonumber\\&&\leq\Pr\Bigg(\left\{\left(\mathbf x, \hat {\mathbf x}_l\right) \in \mathbb{A}_{\epsilon^\prime}^{\ast\left(n\right)}\big(p_{X,\hat X}\big)\quad \forall l\right\}\cap  \left\{\mathbb L \quad \text{is not empty}\right\}
		% \left\{\exists \bar{l}:\bigg( \hat {\mathbf x}_{\bar{l}},\mathbf Y^{\left(\bar{l}\right)}\left(w^{\left(\bar{l}\right)}\right)\bigg) \in \mathbb{A}_{\epsilon}^{\ast\left(n\right)}\big(p_{\hat X,Y}\big)\right\}
		\nonumber\\&&\quad \quad \quad \cap \left\{\exists l\in \mathbb L:\bigg({\mathbf x},\mathbf Y^{\left(l\right)}\left(w^{\left(l\right)}\right)\bigg) \notin \mathbb{A}_{\epsilon}^{\ast\left(n\right)}\big(p_{ X,Y}\big)\right\}  \Bigg)\nonumber\end{IEEEeqnarray}\begin{IEEEeqnarray}{rCl}&&\leq \Pr\Bigg(\exists l:\left\{\left(\mathbf x, \hat {\mathbf x}_l\right) \in \mathbb{A}_{\epsilon^\prime}^{\ast\left(n\right)}\big(p_{X,\hat X}\big)\right\}\nonumber\\&& \quad \quad \quad \quad \quad \cap  \left\{\bigg( \hat {\mathbf x}_{l},\mathbf Y^{\left(l\right)}\left(w^{\left(l\right)}\right)\bigg) \in \mathbb{A}_{\epsilon}^{\ast\left(n\right)}\big(p_{\hat X,Y}\big)\right\}
		% \left\{\bigg( \hat {\mathbf x}_{\bar{l}},\mathbf Y^{\left(\bar{l}\right)}\left(w^{\left(\bar{l}\right)}\right)\bigg) \in \mathbb{A}_{\epsilon}^{\ast\left(n\right)}\big(p_{\hat X,Y}\big)\right\}
		\nonumber\end{IEEEeqnarray}\begin{IEEEeqnarray}{rCl}&&\quad \quad \quad \quad \quad \quad  \cap \left\{\bigg({\mathbf x},\mathbf Y^{\left(l\right)}\left(w^{\left(l\right)}\right)\bigg) \notin \mathbb{A}_{\epsilon}^{\ast\left(n\right)}\big(p_{ X,Y}\big)\right\}  \Bigg)\nonumber\\&&\leq \Pr\Bigg(\exists l: \left\{\left(\mathbf x, \hat {\mathbf x}_l\right) \in \mathbb{A}_{\epsilon^\prime}^{\ast\left(n\right)}\big(p_{X,\hat X}\big)\right\}
		% \left\{\bigg( \hat {\mathbf x}_{\bar{l}},\mathbf Y^{\left(\bar{l}\right)}\left(w^{\left(\bar{l}\right)}\right)\bigg) \in \mathbb{A}_{\epsilon}^{\ast\left(n\right)}\big(p_{\hat X,Y}\big)\right\}
		\nonumber\\&& \quad \quad \quad  \cap \left\{\bigg({\mathbf x},\hat{\mathbf x}_l,\mathbf Y^{\left(l\right)}\left(w^{\left(l\right)}\right)\bigg) \notin \mathbb{A}_{\epsilon}^{\ast\left(n\right)}\big(p_{ X,\hat{X},Y}\big)\right\}  \Bigg)\IEEEeqnarraynumspace \label{caseb6} \\
			&&\leq \sum_{l=1}^{L}\underbrace{\Pr\bigg(\left\{\left(\mathbf x, \hat {\mathbf x}_l\right) \in \mathbb{A}_{\epsilon^\prime}^{\ast\left(n\right)}\big(p_{X,\hat X}\big)\right\}\bigg)}_{\leq 1}\nonumber\\&& \quad \quad \quad\Pr\Bigg( \bigg({\mathbf x},\hat{\mathbf x}_l,\mathbf Y^{\left(l\right)}\left(w^{\left(l\right)}\right)\bigg) \notin \mathbb{A}_{\epsilon}^{\ast\left(n\right)}\big(p_{ X,\hat{X},Y}\big)\nonumber\\&&\quad \quad \quad \quad \quad \quad  \Big|
			\left(\mathbf x, \hat {\mathbf x}_l\right) \in \mathbb{A}_{\epsilon^\prime}^{\ast\left(n\right)}\big(p_{X,\hat X}\big)
			% \left\{\bigg( \hat {\mathbf x}_{\bar{l}},\mathbf Y^{\left(\bar{l}\right)}\left(w^{\left(\bar{l}\right)}\right)\bigg) \in \mathbb{A}_{\epsilon}^{\ast\left(n\right)}\big(p_{\hat X,Y}\big)\right\}
			\Bigg)\nonumber\\
		%&&\leq \sum_{l=1}^{L}  \Pr\Bigg(\left\{\bigg( \hat {\mathbf x}_{l},\mathbf Y^{\left(l\right)}\left(w^{\left(l\right)}\right)\bigg) \in \mathbb{A}_{\epsilon}^{\ast\left(n\right)}\big(p_{\hat X,Y}\big)\right\}
		% \left\{\bigg( \hat {\mathbf x}_{\bar{l}},\mathbf Y^{\left(\bar{l}\right)}\left(w^{\left(\bar{l}\right)}\right)\bigg) \in \mathbb{A}_{\epsilon}^{\ast\left(n\right)}\big(p_{\hat X,Y}\big)\right\}
		%\\&&\quad  \cap \left\{\bigg({\mathbf x},\hat{\mathbf x}_l,\mathbf Y^{\left(l\right)}\left(w^{\left(l\right)}\right)\bigg) \notin \mathbb{A}_{\epsilon}^{\ast\left(n\right)}\big(p_{ X,\hat{X},Y}\big)\right\}  \Bigg)
		&&\leq \sum_{l=1}^{L}  \Pr\Bigg( \bigg({\mathbf x},\hat{\mathbf x}_l,\mathbf Y^{\left(l\right)}\left(w^{\left(l\right)}\right)\bigg) \notin \mathbb{A}_{\epsilon}^{\ast\left(n\right)}\big(p_{ X,\hat{X},Y}\big)\nonumber\\&&\quad \quad\quad \quad \quad \quad \quad  \Big|
		\left(\mathbf x, \hat {\mathbf x}_l\right) \in \mathbb{A}_{\epsilon^\prime}^{\ast\left(n\right)}\big(p_{X,\hat X}\big)
		% \left\{\bigg( \hat {\mathbf x}_{\bar{l}},\mathbf Y^{\left(\bar{l}\right)}\left(w^{\left(\bar{l}\right)}\right)\bigg) \in \mathbb{A}_{\epsilon}^{\ast\left(n\right)}\big(p_{\hat X,Y}\big)\right\}
		\Bigg)\nonumber\\&&= \sum_{l=1}^{L} 1-\Pr\Bigg( \bigg({\mathbf x},\hat{\mathbf x}_l,\mathbf Y^{\left(l\right)}\left(w^{\left(l\right)}\right)\bigg) \in \mathbb{A}_{\epsilon}^{\ast\left(n\right)}\big(p_{ X,\hat{X},Y}\big)\nonumber\end{IEEEeqnarray} 
	\begin{IEEEeqnarray}{rCl}&&\quad \quad \quad \quad \quad \quad \Big|
	\left(\mathbf x, \hat {\mathbf x}_l\right) \in \mathbb{A}_{\epsilon^\prime}^{\ast\left(n\right)}\big(p_{X,\hat X}\big)
		% \left\{\bigg( \hat {\mathbf x}_{\bar{l}},\mathbf Y^{\left(\bar{l}\right)}\left(w^{\left(\bar{l}\right)}\right)\bigg) \in \mathbb{A}_{\epsilon}^{\ast\left(n\right)}\big(p_{\hat X,Y}\big)\right\}
		\Bigg)\nonumber\\&&\leq  \sum_{l=1}^{L}\delta_{t}\left(n,\epsilon/2,\mathbb{X}\times \mathbb X\times \mathbb Y\right) \nonumber,\end{IEEEeqnarray}
	where \eqref{caseb6} follows because, due to the fact that (strong) joint typicality implies pairwise typicality, we enlarge the set and the last step follows from Lemma \ref{lem:markov} (in Appendix B).
	% Lemma \ref{lem:markov} (in Appendix) gives us that  $\Pr\left(\text {Case c}\right)\leq \delta_t\left(n,\epsilon/2,\mathbb X \times \mathbb { X}\times \mathbb Y\right)$.
	%To bound the expected distortion in Case (c), we note that if $\left(\mathbf x,\hat{\mathbf Y}^{\left(1\right)},\hat{\mathbf Y}^{\left(2\right)},\hat{\mathbf Y}^{\left(12\right)}\right)$
	%is jointly typical, then, each pair $\left(\mathbf x,\hat{\mathbf Y}^{\left(1\right)}\right)$,$\left(\mathbf x,\hat{\mathbf Y}^{\left(2\right)}\right)$,$\left(\mathbf x,\hat{\mathbf Y}^{\left(12\right)}\right)$ is jointly typical so \begin{IEEEeqnarray*}{rCl}&&\|P_{x^n,y^n}\left(x,y\right)-p_{0}\left(x\right)p\left(y|x\right)\|_{TV}\nonumber\\&& \leq\|P_{x^n,y^n}\left(x,y\right)-p_{0}\left(x\right)p_{\hat{Y}^{\left(i\right)}|X}\left(y|x\right)\|_{TV}\nonumber\\&& +\|p_{0}\left(x\right)p_{\hat{Y}^{\left(i\right)}|X}\left(y|x\right)-p_{0}\left(x\right)p\left(y|x\right)\|_{TV}\label{eq:trianglein}\leq\frac{\epsilon}{2}+\Delta_{i},\nonumber
	%\end{IEEEeqnarray*}
	%by choosing $p_{\hat{Y}^{\left(i\right)}|X}\left(y|x\right)$ such that $p_{0}\left(x\right)p_{\hat{Y}^{\left(i\right)}|X}\left(y|x\right)\in N_{\Delta_{i}}$ for $i=1,2,12$.
	
\end{itemize}
%\begin{equation*}\hat{\mathbf Y}^{\left(2\right)}\left(w_2\right) \quad  \text{if only $w_2$ is received,}\end{equation*}
%\begin{equation*}\hat{\mathbf Y}^{\left(12\right)}\left(w_1,w_2\right)  \quad  \text{if both $\left(w_1,w_2\right)$ are received.} \end{equation*}
%\textbf{Converse}.	
\end{IEEEproof}
\begin{IEEEproof}[Proof of Theorem 2]
\begin{itemize}
	\item{\textit{Setup}:} We assume that $\epsilon_{\text{ag}}>0$ is given. We fix some rates per agent $R_{\text{ag}}$ and $R^{\prime}_{\text{ag}}$, some blocklength $n$, some $\epsilon>0$, $\epsilon_0>0$
	%$\epsilon_1,\dots,\epsilon_L>0$ 
	and for every PMF $p_{X,\hat X,Y}=p_Xp_{\hat X|X}p_{Y|\hat X}$ compute the marginal $p_{Y}$.	\item{\textit{Codebook design}:} Generate $\lfloor e^{n\left(R_{\text{ag}}+\epsilon_{\text{ag}}\right)} \rfloor \lceil  e^{n\left(R^{\prime}_{\text{ag}}-\epsilon_0\right)} \rceil$ length-$n$ codewords $\mathbf Y^{\left(l\right)}\left(w^{\left(l\right)}, v^{\left(l\right)}\right), w^{\left(l\right)}=1,\dots, \lfloor e^{n\left(R_{\text{ag}}+\epsilon_{\text{ag}}\right)} \rfloor,v^{\left(l\right)}=1,\dots, \lceil e^{n\left(R^{\prime}_{\text{ag}}-\epsilon_0\right)} \rceil$, by choosing each of the $n\lfloor e^{n\left(R_{\text{ag}}+\epsilon_{\text{ag}}\right)} \rfloor \lceil  e^{n\left(R^{\prime}_{\text{ag}}-\epsilon_0\right)} \rceil $ symbols $Y^{\left(l\right)}_k\left(w^{\left(l\right)},v^{\left(l\right)}\right)$ independently at random according to $p_{Y}$ for $l=1,\dots,L$.
	\item{\textit{Encoder Design}:} For given sequences $\mathbf{x},\hat{\mathbf{x}}_1,\dots,\hat{\mathbf{x}}_L$, the $l$-th encoder tries to find a pair $\left(w^{\left(l\right)},v^{\left(l\right)}\right)$ such that 
	\begin{multline}
	\bigg( \hat {\mathbf x}_l,\mathbf Y^{\left(l\right)}\left(w^{\left(l\right)},v^{\left(l\right)}\right)\bigg)
	\in \mathbb{A}_{\epsilon}^{\ast\left(n\right)}\big(p_{\hat{X},Y}\big).\label{eq:jtypicality2}
	\end{multline}
	If it finds several possible choices, they pick the first. If it finds none, it declares an error. The $l$-th encoder puts out $w^{\left(l\right)}$.
	\item{\textit{Decoder Design}:} The decoder $y^n$ based on the bin numbers $\Big(w^{\left(1\right)},\dots,w^{\left(L\right)}\Big)$ that receives, it tries to find a tuple $\Big(v^{\left(1\right)},\dots,v^{\left(L\right)}\Big)$ and an $\mathbf x$ such that  
	$\mathbf Y^{nL}= \bigg(\mathbf Y^{\left(1\right)}\left(w^{\left(1\right)},v^{\left(1\right)}\right),\dots,\mathbf Y^{\left(L\right)}\left(w^{\left(L\right)},v^{\left(L\right)}\right)\bigg)$ and $\mathbf x^{nL}=\left(\underbrace{\mathbf x,\dots,\mathbf x}_{L \quad \text{times}}\right)$ to be jointly typical i.e.,
	\begin{IEEEeqnarray}{r}\left(\mathbf x^{nL},\mathbf Y^{nL}\right)\in \mathbb{A}_{\epsilon}^{\ast\left(nL\right)}\big(p_{X,Y}\big).\label{eq:jtypicality3}
	\end{IEEEeqnarray}
	If it finds more than one $\big(v^{\left(1\right)},\dots,v^{\left(L\right)}\big)$ or none, it declares an error. Otherwise, it chooses some $j$ and puts out $\mathbf Y^{\left(j\right)}\left(w^{\left(j\right)},v^{\left(j\right)}\right)$.
	% from the class of $\left(\mathbf Y^{nL},\mathbf x^{nL}\right)$ which satisfy \eqref{eq:Ltypicality}  and 
	%based on the property that $P_{ Y^{nL}, x^{nL}}=\frac{1}{L}\sum_{i=1}^{L}{P_{ Y^{\left(i\right)L}, x^{\left(i\right)L}}}$, the decoder finds always a pair $\Big(\mathbf Y^{\left(j\right)L}, \mathbf x\Big)$ such that $\Big(\mathbf Y^{\left(j\right)L}, \mathbf x\Big)\in \mathbb{A}_{\epsilon}^{\ast\left(n\right)}\big(p_{X,Y}\big)$.
	\item{\textit{Performance Analysis}:} We define $\epsilon^\prime=\frac{\epsilon}{2|\mathbb X|}$ %We define $\epsilon^\prime\triangleq\frac{\epsilon}{2|\mathbb Y|^3}$
	and partition the error space into four disjoint cases:
	%The source sequence $\mathbf x$ is not typical:
	(a) $\left(\mathbf x, \hat {\mathbf x}_l\right) \notin \mathbb{A}_{\epsilon^\prime}^{\ast\left(n\right)}\big(p_{X,\hat X}\big)$ for some $l$
	%in which case we for sure cannot find a pair $\left(w_1,w_2\right)$ such that \eqref{jtypicality} is satisfied
	(b) $\left(\mathbf x, \hat {\mathbf x}_l\right) \in \mathbb{A}_{\epsilon^\prime}^{\ast\left(n\right)}\big(p_{X,\hat X}\big)$ for every $l$ but at least one encoder declares an error %i.e. \eqref{jtypicality} is not satisfied for every l=1,\dots,L.
	(c)  $\left(\mathbf x, \hat {\mathbf x}_l\right) \in \mathbb{A}_{\epsilon^\prime}^{\ast\left(n\right)}\big(p_{X,\hat X}\big)$ for every $l$, all encoders do not declare an error but the decoder finds none $\big(v^{\left(1\right)},\dots,v^{\left(L\right)}\big)$ (event $C_{\text{a}}$) or more than one (event $C_{\text{b}}$)  
	(d)  $\left(\mathbf x, \hat {\mathbf x}_l\right) \in \mathbb{A}_{\epsilon^\prime}^{\ast\left(n\right)}\big(p_{X,\hat X}\big)$ for every $l$, all encoders do not declare an error and the decoder finds exactly one $\big(v^{\left(1\right)},\dots,v^{\left(L\right)}\big)$ but \eqref{eq:empcord1} is not satisfied.
	%By the Union Bound and Lemma \ref{lem:TA} (in Appendix)
	By the Union Bound and Lemma 5 (in Appendix 5), we get
	%we can bound the probability of Case (a) as
	$\Pr\left(\text{Case a}\right)\leq \sum_{l=1}^{L}\delta_{t}\left(n,\epsilon^\prime,\mathbb{X\times X}\right)$.
	Easily, it follows that  $\Pr\left(\text{Case b}\right)\leq \sum_{l=1}^{L}\exp\Big(-e^{n\left(R_{\text{ag}}+R^\prime_{\text{ag}}-I\left(\hat X;Y\right)+\epsilon_{\text{ag}}-\delta\right)}\Big)$, where $\delta$ accounts for the rounding mistake and includes the $2\epsilon_m$-term and the $\left(1-\delta_t\right)$-factor. Hence, we see that as long as $n$ is large enough, $R_{\text{ag}}+R^\prime_{\text{ag}}\geq I\left(\hat X;Y\right)$ and $\epsilon$ small enough such that $\delta<\epsilon_{\text{ag}}$, the probability $\Pr\left(\text {Case b}\right)$ tends to zero double-exponentially fast in $n$.
	In case (c), we have $\Pr\left(\text{Case c}\right)
	=\Pr\left(C_{\text{a}}\cup C_{\text{b}}\right)
	%\cap \left(\text{Case} \quad b\right)^c\cap \left(\text{Case} \quad a\right)^c\big)\leq \Pr\left( C_a\cup C_b\right)= 
	=\Pr\left( C_{\text{a}}\right)+ \Pr\left( C_{\text{b}}-C_{\text{a}} \right)$.
	By \eqref{eq:jtypicality2}, the fact that  $\left(\mathbf x, \hat {\mathbf x_l}\right) \in \mathbb{A}_{\epsilon^\prime}^{\ast\left(n\right)}\big(p_{X,\hat X}\big)$ for every $l$ and the simple properties 	$P_{\hat{x}^{nL},Y^{\left(nL\right)}}=\frac{1}{L}\sum_{l=1}^{L}{P_{\hat{x}_{l},Y^{\left(l\right)}}}$, $P_{x^{nL},\hat{x}^{nL}}=\frac{1}{L}\sum_{l=1}^{L}{P_{x,\hat{x}_{l}}}$, it follows that $\left(\hat{\mathbf{x}}^{nL},\mathbf Y^{nL}\right)\in \mathbb{A}_{\epsilon}^{\ast\left(nL\right)}\big(p_{\hat{X},Y}\big)$ and $\left(\mathbf{x}^{nL},\hat{\mathbf{x}}^{nL}\right)\in \mathbb{A}_{\epsilon}^{\ast\left(nL\right)}\big(p_{X,\hat{X}}\big)$  where $ \hat{\mathbf{x}}^{nL}=\left(\hat {\mathbf{ x}}_1,\dots,\hat {\mathbf{ x}}_L\right)$. Lemma \ref{lem:markov} (in Appendix B) gives us that  $\Pr\left(C_{\text{a}}\right)\leq\delta_t\left(nL,\epsilon/2,\mathbb X \times \mathbb { X}\times \mathbb Y\right)$. 
	We proceed with the event $C_{\text{b}}-C_{\text{a}}$. The cardinality of the set $\tilde{\mathbb Y}^{nL}\subseteq\mathbb Y^{nL}$ of the codewords which satisfy \eqref{eq:jtypicality3} is bounded as
	$|\tilde{\mathbb Y}^{nL}|\leq|\mathbb{A}_{\epsilon}^{\ast\left(n\right)}\big(p_{X}\big)|\max_{\substack{\mathbf x^{nL}\in \mathbb{A}_{\epsilon}^{\ast\left(nL\right)}\big(p_{X}\big)}}\big|\mathbb{A}_{\epsilon}^{\ast\left(nL\right)}\big(p_{X,Y}|\mathbf x^{nL}\big)\big|\leq e^{n\big(H\left(X\right)+\epsilon_m\big)}e^{nL\big(H\left(Y|X\right)+\epsilon_m\big)}\leq e^{nL\big(H\left(Y|X\right)+4\epsilon_m\big)}$, %\label{eq:cb1}
	where the second inequality follows from Lemma \ref{lem:TA} and Lemma \ref{lem:TB} (in Appendix B) and the last inequality is true for $L\geq H\left(X\right)/{\epsilon_m}$. The probability for each element of this set to be chosen to a specific bin-tuple  $\Big(w^{\left(1\right)},\dots,w^{\left(L\right)}\Big)$ is due to Lemma \ref{lem:TA} (in Appendix B) at most 
	$e^{-nL\big(H\left(Y\right)-\epsilon_m\big)}e^{nL\left(R^{\prime}_{\text{ag}}-\epsilon_0\right)}=e^{-nL\big(H\left(Y\right)-R^{\prime}_{\text{ag}}-\epsilon_m+\epsilon_0\big)}$. Hence, combining these two give us that $\Pr\left( C_{\text{b}}-C_{\text{a}}\right)\leq e^{-nL\big(I(X;Y)-R^{\prime}_{\text{ag}}-5\epsilon_m+\epsilon_0\big)}$. 
	Therefore, we see that as long as $n$ is large enough, $L\geq H\left(X\right)/{\epsilon_m}$, $R^{\prime}_{\text{ag}}\leq I\left( X;Y\right)$ and $\epsilon$ small enough such that $\epsilon_m<\epsilon_0/5$, the probability $\Pr\left(\text {Case c}\right)$ tends to zero exponentially. Lemma \ref{lem:markov} (in Appendix B) guarantees again that $\Pr\left(\text {Case d}\right)$ decays. So, collecting all the cases together gives us the desired result.
\end{itemize}	
\end{IEEEproof}
\section*{\quad \quad \quad \quad \quad  Appendix B \newline Typical Sets}

\begin{lemma}[\cite{moser:2019}] $\forall \theta>0, \quad \forall \xi\leq 1:
\left(1-\xi\right)^\theta\leq e^{\theta\xi}$. \label{lem:expineq}
\end{lemma}
 \begin{definition}[Strongly $\epsilon$-typical sets \cite{moser:2019}]
	%Fix an $\epsilon>0$, a PMF $p_{X,Y}\left(x,y\right)$, and a blocklength n. The strongly typical set $\mathbb{A}_{\epsilon}^{\ast\left(n\right)}\left(p_{X,Y}\right)$ with respect to the PMF $p_{X,Y}\left(x,y\right)$ is defined as 
	%\begin{align*}
	%\bigg\{\mathbf x\in \mathbb X^n:&\quad |P_{\mathbf x}\left(a\right)-Q_X\left(a\right)|<\frac{\epsilon}{|\mathbb X|},\quad \forall a\in \mathbb X \quad \text{and}\\ &\quad|P_{\mathbf x}\left(a\right)|=0 \quad \forall a\in \mathbb X \quad \text{with} \quad Q_X\left(a\right)=0 \bigg\}.
	%\end{align*}
	\begin{multline*}
	\mathbb{A}_{\epsilon}^{\ast\left(n\right)}\left(p_{X,Y}\right)\triangleq\\
	\left.
	\left\{ \,
	\begin{IEEEeqnarraybox}[
	\IEEEeqnarraystrutmode
	\IEEEeqnarraystrutsizeadd{2pt}
	{2pt}][c]{l}
	\left(\mathbf x,\mathbf y\right)\in \mathbb X^n\times \mathbb Y^n:\\
	|P_{\mathbf x,\mathbf y}\left(a,b\right)-p_{X,Y}\left(a,b\right)|<\frac{\epsilon}{|\mathbb X||\mathbb Y|},\forall \left(a,b\right)\in \mathbb X\times \mathbb Y, %\\\text{and} \quad P_{\mathbf x,\mathbf y}\left(a,b\right)=0 ,\quad \forall \left(a,b\right)\in \mathbb X\times \mathbb Y \quad \text{with}\\ p_{X,Y}\left(a,b\right)=0
	% \quad  \text{and}\quad |P_{\mathbf x,\mathbf y}\left(a,b\right)|=0\\ \forall \left(a,b\right)\in \mathbb X\times \mathbb Y \quad \text{with} \quad p_{X,Y}\left(a,b\right)=0
	\end{IEEEeqnarraybox}\right\},
	\right.
	%\label{eq:example_left_right2}.
	\label{def:typsets}
	\end{multline*}

%\begin{definition}[Conditionally $\epsilon$-typical set  \cite{moser:2019}]
%	For some joint PMF $p_{X,Y}$ with marginal $p_X$ and for some given sequence $\mathbf x \in \mathbb{A}_{\epsilon}^{\ast\left(n\right)}\left(p_X\right)$, we define the conditionally strongly typical set with respect to $p_{X,Y}$ as 
	\begin{IEEEeqnarray*}{rCl}&&\mathbb{A}_{\epsilon}^{\ast\left(n\right)}\big(p_{X,Y}|\mathbf x\big)\triangleq \left\{\mathbf y\in \mathbb Y^n:\left(\mathbf x,\mathbf y\right)\in \mathbb{A}_{\epsilon}^{\ast\left(n\right)}\big(p_{X,Y}\big) \right\}.
	\end{IEEEeqnarray*}
	\label{def:ctypsets}
\end{definition}
%\begin{lemma}[\cite{moser:2019}]The event $\Big\{\left(\mathbf X,\mathbf Y\right)\in \mathbb{A}_{\epsilon}^{\ast\left(n\right)}\left(p_{X,Y}\right) \Big\}$
	%is equivalent to the event 
	%\begin{IEEEeqnarray*}{rCl}&&
		%\left\{\mathbf X\in \mathbb{A}_{\epsilon}^{\ast\left(n\right)}\left(p_{X}\right) \right\}\cap \left\{\mathbf Y\in \mathbb{A}_{\epsilon}^{\ast\left(n\right)}\big(p_{X,Y}|\mathbf X\big)\right\}.
	%\end{IEEEeqnarray*}\label{lem:chtypsets}
%\end{lemma}
	\begin{lemma}[\cite{moser:2019}] $\Big\{\left(\mathbf X,\mathbf Y\right)\in \mathbb{A}_{\epsilon}^{\ast\left(n\right)}\left(p_{X,Y}\right) \Big\} \iff
 \left\{\mathbf X\in \mathbb{A}_{\epsilon}^{\ast\left(n\right)}\left(p_{X}\right) \right\}\cap \left\{\mathbf Y\in \mathbb{A}_{\epsilon}^{\ast\left(n\right)}\big(p_{X,Y}|\mathbf X\big)\right\}$.
\label{lem:chtypsets}
\end{lemma}
\begin{definition}[$\epsilon_m,\delta_t$,  \cite{moser:2019}]
$\epsilon_m\big(p_{X,Y}\left(x,y\right)\big)\triangleq-\epsilon\log\left(p_{X,Y}^{\min}\right)$, $\delta_{t}\left(n,\epsilon,\mathbb{X}\times \mathbb{Y} \right)\triangleq \left(n+1\right)^{|\mathbb X||\mathbb Y|}e^{-n\frac{\epsilon^2}{2|\mathbb X|^2|\mathbb Y|^2}\log e}$,
	where $p_{X,Y}^{\min}$ is the smallest value of $p_{X,Y}\left(x,y\right)$.
	\label{def:epsilondelta}
\end{definition}
\begin{lemma}[\cite{moser:2019}] Let $\left(\mathbf x,\mathbf y\right) \in \mathbb{A}_{\epsilon}^{\ast\left(n\right)}\left(p_{X,Y}\right)$. Then,
$e^{-n\Big(H\left(X,Y\right)+\epsilon_m\big(p_{X,Y}\left(x,y\right)\big)\Big)}< p_{X,Y}^n\left(\mathbf x,\mathbf y\right)< e^{-n\Big(H\left(X,Y\right)-\epsilon_m\big(p_{X,Y}\left(x,y\right)\big)\Big)}$.	Moreover, 
$1-\delta_{t}\left(n,\epsilon,\mathbb{X}\times \mathbb{Y}\right)\leq\Pr\left[\left(\mathbf x, \mathbf y\right) \in \mathbb{A}_{\epsilon}^{\ast\left(n\right)}\left(p_{X,Y}\right)\right]\leq 1$ and
$|\mathbb{A}_{\epsilon}^{\ast\left(n\right)}\big(p_{X,Y}\big)|< e^{n\Big(H\left(X,Y\right)+\epsilon_m\big(p_{X,Y}\left(x,y\right)\big)\Big)}$. 
	\label{lem:TA}
\end{lemma}	
\begin{lemma}[\cite{moser:2019}]
$|\mathbb{A}_{\epsilon}^{\ast\left(n\right)}\big(p_{X,Y}|\mathbf x\big)|< e^{n\Big(H\left(Y|X\right)+\epsilon_m\big(p_{X,Y}\left(x,y\right)\big)\Big)}$.
%	If also $\mathbf x \in\mathbb{A}_{\frac{\epsilon}{2|\mathbb Y|}}^{\ast\left(n\right)}\left(p_{X}\right)$, then, we can obtain 
%	\begin{IEEEeqnarray*}{rCl}&&|\mathbb{A}_{\epsilon}^{\ast\left(n\right)}\big(p_{X,Y}|\mathbf x\big)|> \left(1-\delta_{t}\big(n,\frac{\epsilon}{2},\mathbb{X}\times \mathbb{Y} \big)\right)\\&&\cdot e^{n\Big(H\left(Y|X\right)-\epsilon_m\big(p_{X,Y}\left(x,y\right)\big)\Big)}.\end{IEEEeqnarray*}
	\label{lem:TB}
\end{lemma}
\begin{lemma}[Markov Lemma,\cite{moser:2019}] Let a PMF  $p_{U,V,W}\left(u,v,w\right)$ with a Markov stracture 
$U-V-W$
%\begin{IEEEeqnarray*}{r}
%p_{U,V,W}=p_{U}p_{V|U}p_{W|V},
%\end{IEEEeqnarray*}
and let $\left(\mathbf u,\mathbf v\right)\in \mathbb{A}_{\epsilon ^\prime}^{\ast\left(n\right)}\big(p_{U,V}\big)$ with
$\epsilon ^ \prime \triangleq \frac{\epsilon}{2|\mathbb W|}$.
Then, 
$p_{W|V}^n\left(\mathbb{A}_{\epsilon}^{\ast\left(n\right)}\big(p_{U,V,W}|\mathbf u,\mathbf v\big)|\mathbf v\right)\geq 1-\delta_t\left(n,\epsilon/2,\mathbb U \times \mathbb V \times \mathbb W \right)$.
\label{lem:markov}
\end{lemma}
\begin{lemma}[\cite{moser:2019}] Let $p_{X,Y}\left(x,y\right)$ be a joint PMF with marginals $p_X\left(x\right), p_Y\left(y\right)$. Let $\left(\mathbf x,\mathbf y\right)$ be generated:$\left\{\left(x_k,y_k\right)\right\}_{k=1}^n \text{IID} \sim p_X\left(x\right)p_Y\left(y\right).$
	%	\begin{IEEEeqnarray*}{rCl}&&\left\{\left(x_k,y_k\right)\right\}_{k=1}^n \text{IID} \sim p_X\left(x\right)p_Y\left(y\right).\end{IEEEeqnarray*}
%Then,
%	%\begin{IEEEeqnarray*}{rCl}&&\big(1-\delta_{t}\left(n,\epsilon,\mathbb{X}\times \mathbb{Y} \right)\big)e^{-n\big(I\left(X;Y\right)+\epsilon_2\big)}\\&&<\Pr\left[\left(\mathbf x,\mathbf y\right) \in \mathbb{A}_{\epsilon}^{\ast\left(n\right)}\left(p_{X,Y}\right)\right]<e^{-n\big(I\left(X;Y\right)-\epsilon_2\big)},\end{IEEEeqnarray*} where 
%	$\epsilon_2\triangleq \epsilon_m\big(p_{X,Y}\left(x,y\right)\big)+\epsilon_m\big(p_X\left(x\right)\big)+\epsilon_m\big(p_Y\left(y\right)\big)\leq 3\epsilon_m\big(p_{X,Y}\left(x,y\right)\big).$ Moreover,
%	\begin{IEEEeqnarray*}{rCl}&&
	%	\Pr\left[\mathbf Y\in \mathbb{A}_{\epsilon}^{\ast\left(n\right)}\big(p_{X,Y}|\mathbf x\big)\right]<e^{-n\big(I\left(X;Y\right)-\epsilon_3\big)},
%	\end{IEEEeqnarray*}
If $\mathbf x \in \mathbb{A}_{\frac{\epsilon}{2|\mathbb Y|}}^{\ast\left(n\right)}\left(p_{X}\right)$, then,  we obtain 
$\Pr\left[\mathbf Y\in \mathbb{A}_{\epsilon}^{\ast\left(n\right)}\big(p_{X,Y}|\mathbf x\big)\right]>\big(1-\delta_t\left(n,\frac{\epsilon}{2},\mathbb X\times \mathbb Y\right)\big) e^{-n\big(I\left(X;Y\right)+\epsilon_3\big)}$
where $\epsilon_3\triangleq \epsilon_m\big(p_{X,Y}\left(x,y\right)\big)+\epsilon_m\big(p_Y\left(y\right)\big)\leq 2\epsilon_m\big(p_{X,Y}\left(x,y\right)\big)$.
\label{lem:TC}
\end{lemma}

%\begin{lemma}[\cite{moser:2019}]
%For every $\mathbf y\in \mathbb{A}_{\epsilon}^{\ast\left(n\right)}\left(p_{X,Y}|\mathbf x\right)$ we obtain 
%	\begin{IEEEeqnarray*}{rCl}&&
%	e^{-n\Big(H\left(Y|X\right)+\epsilon_m\big(p_{X,Y}\left(x,y\right)\big)\Big)}< p_{Y|X}^n\left(\mathbf y|\mathbf x\right)\\&&< e^{-n\Big(H\left(Y|X\right)-\epsilon_m\big(p_{X,Y}\left(x,y\right)\big)\Big)}.
%	\end{IEEEeqnarray*}
%	The size of the conditionally strongly typical set is bounded as 
%	\begin{IEEEeqnarray*}{rCl}&&|\mathbb{A}_{\epsilon}^{\ast\left(n\right)}\big(p_{X,Y}|\mathbf x\big)|< e^{n\Big(H\left(Y|X\right)+\epsilon_m\big(p_{X,Y}\left(x,y\right)\big)\Big)}.\end{IEEEeqnarray*}
%	If also $\mathbf x \in\mathbb{A}_{\frac{\epsilon}{2|\mathbb Y|}}^{\ast\left(n\right)}\left(p_{X}\right)$, then, we can obtain 
%	\begin{IEEEeqnarray*}{rCl}&&|\mathbb{A}_{\epsilon}^{\ast\left(n\right)}\big(p_{X,Y}|\mathbf x\big)|> \left(1-\delta_{t}\big(n,\frac{\epsilon}{2},\mathbb{X}\times \mathbb{Y} \big)\right)\\&&\cdot e^{n\Big(H\left(Y|X\right)-\epsilon_m\big(p_{X,Y}\left(x,y\right)\big)\Big)}.\end{IEEEeqnarray*}
%	\label{lem:TB}
%\end{lemma}
% Bibliography %=================================================================================

\bibliographystyle{IEEEtran}
\bibliography{string,references}

\end{document}